\documentclass[aps,prl,showpacs,groupedaddress,twocolumn]{revtex4}
\usepackage{graphicx}
\usepackage{latexsym}
\usepackage{stmaryrd}

\begin{document}

\title{Topological insulator: a new quantized spin Hall resistance robust to dephasing}

\author{Hua Jiang$^{1}$, Shuguang Cheng$^{1}$,  Qing-feng Sun$^{1,\ast}$, and X. C. Xie$^{2,1}$}

\address {\ $^1$Beijing National Lab for Condensed Matter Physics and
Institute of Physics, Chinese Academy of Sciences, Beijing 100190,
China;\\
$^2$Department of Physics, Oklahoma State University, Stillwater,
Oklahoma 74078 }

\date{\today}

\begin{abstract}
The influence of dephasing on the quantum spin Hall effect (QSHE) is
studied. In the absence of dephasing, the longitudinal resistance in
a QSHE system exhibits the quantum plateaus. We find that these
quantum plateaus are robust against the normal dephasing but fragile
with the spin dephasing. Thus, these quantum plateaus only survive
in mesoscopic samples. Moreover, the longitudinal resistance
increases linearly with the sample length but is insensitive to the
sample width. These characters are in excellent agreement with the
recent experimental results [science {\bf 318}, 766 (2007)]. In
addition, we define a new spin Hall resistance that also exhibits
quantum plateaus. In particular, these plateaus are robust against
any type of dephasing and therefore, survive in macroscopic samples
and better reflect the topological nature of QSHE.

\end{abstract}

\maketitle

Recently, the quantum spin Hall effect (QSHE), existed in a new
quantum state of matter with a non-trivial topological property, has
generated great interest.\cite{ref1} QSHE occurs in the topological
insulator with a bulk energy gap and two helical edge states
crossing inside the gap. This energy-band structure guarantees that
the carriers only flow along the boundary and that carriers with
opposite spin-polarizations move in opposite directions on a given
edge. The other key ingredient for QSHE is the presence of the
spin-orbit interaction (SOI). When electrons move under an electric
field, the SOI drives the electrons with opposite spins to deflect
to the opposite transverse boundaries, and the special energy-band
structure leads to the quantum spin Hall conductance.\cite{ref1} The
existence of QSHE was first proposed in a graphene film in which the
SOI opens a bandgap around the Dirac-points and establishes the edge
states.\cite{ref3,ref4}. Soon afterwards, QSHE was predicted to
exist in some other two- or three-dimensional
systems.\cite{ref5,ref6,ref7,ref8} In particular, Bernevig {\sl et
al.} recently found that CdTe/HgTe/CdTe quantum well has an
"inverted" type energy-band structure with proper well
thicknesses\cite{ref8} where QSHE naturally exists. Soon after this
work, QSHE was successfully realized in an
experiment\cite{ref9,ref10} in which a quantized longitudinal
resistance plateau was observed when the sample's electron density
was varied in the absence of a magnetic field.\cite{ref9}

However, in the experiment of Ref.\cite{ref9}, the quantized
longitudinal resistance plateaus could only emerge in mesoscopic
samples. This character is very different from the regular quantum
Hall effect (QHE). In QHE, the Hall resistance plateaus exist in
macroscopic samples, robust against the impurity scattering as well
as the inelastic (dephasing) scattering. This leads some to
speculate that the inelastic scattering which induces phase
relaxation, destroys the quantized plateaus in QSHE\cite{ref1,ref8},
however, there has been no theoretical or experimental investigation
thus far.

In this Letter, we study how QSHE is affected by dephasing. We
mainly focus on two questions: (i) How does dephasing affect the
quantized longitudinal resistance plateau of QSHE samples as
measured in the recent experiment\cite{ref9}; (ii) Is there an
observable physical quantity showing a quantized value in
macroscopic samples, reflecting the topological nature of QSHE?

In a realistic sample, there are in general a number of possible
dephasing processes, but these can be classified into two
categories. In the first kind, the carriers lose only the phase
memory while maintaining the spin memory, such as with the dephasing
processes caused by the electron-electron interaction, the
electron-phonon interaction, {\it etc}, these are named normal
dephasing in this paper. In the second kind, the carriers lose both
phase and spin memories, such as with the spin-flip dephasing
processes caused by the magnetic impurities, the nuclear spins, {\it
etc}, named spin dephasing. We consider a six-terminal device (shown
in Fig.1a), as in the experimental set-up\cite{ref9}, and the
dephasing processes are simulated by using the B$\ddot{u}$ttiker's
virtual probes.\cite{ref11} By applying the
Landauer-B$\ddot{u}$ttiker formalism combined with the
non-equilibrium Green function method\cite{ref12,ref13}, the
longitudinal resistance is calculated. The results show that the
longitudinal resistance exhibits the quantum plateaus without
dephasing or with the normal dephasing, but are then destroyed by
the spin dephasing. Thus, these quantum plateaus are only observable
in a mesoscopic sample in which the sample length is smaller than
the spin-dephasing length. Our theoretical results provide a good
understanding of the experimental findings of the dependence of the
longitudinal resistance on temperature, sample length and sample
width. More interestingly, we introduce a novel spin Hall resistance
that also exhibits the quantized plateaus. In particular, these
plateaus survive under both normal and spin dephasings, and are thus
observable in macroscopic samples, similar to the conventional QHE.

In the tight-binding representation, the Hamiltonian of the
six-terminal QSHE device can be written as\cite{ref13,ref14}
\begin{eqnarray}
H&=&- [\sum_{<{\bf i}{\bf j}>\sigma} t e^{i\eta(\sigma)\phi_{{\bf
i}{\bf j}}}c_{{\bf i}\sigma}^{\dagger}c_{{\bf j}\sigma} +H.c.]
  +
[\sum_{{\bf i}k\sigma}\epsilon_{k \sigma} \nonumber \\
 & & a_{{\bf i}k\sigma}^{\dagger} a_{{\bf i}k \sigma}
+(t_{k\sigma}a_{{\bf i}k\sigma}^{\dagger} c_{{\bf i}\sigma}+H.c.) ]
\end{eqnarray}
The first term describes the QSHE system including the central
region and the six terminals. $c_{{\bf i}\sigma}^{\dagger}$
($c_{{\bf i}\sigma}$) is the creation (annihilation) operator of an
electron on the lattice site ${\bf i}$ with spin $\sigma$,
$t=\hbar^2/2m^*a^2$ represents the nearest hopping matrix element
with the lattice constant $a$. Due to the SOI, an extra
spin-dependent phase $\eta(\sigma)\phi_{{\bf i}{\bf j}}$ is added in
the hopping element, with $\eta(\sigma)=1$ and $-1$ for
$\sigma=\uparrow$ and $\downarrow$.\cite{ref15} The summation of
four $\phi_{{\bf i}{\bf j}}$ along each unit satisfies
$\sum_\oblong\phi_{{\bf i}{\bf j}}=e { B_{eff}}a^2/\hbar$, with an
effective magnetic field ${ B_{eff}}$ coming from the SOI. The
second term represents the Hamiltonian of virtual leads and their
couplings to the central sites. Here we assume that the dephasing
only occurs in the central region, and each site ${\bf i}$ in the
central region is attached by a virtual lead. The size of the
central region is $(L+2M)\times W$ as shown in Fig.1a. In
additional, if we take $\eta(\sigma)=1$ in Eq.(1), it describes a
QHE system.\cite{ref16}

Using multi-probe Landauer-B$\ddot{u}$ttiker formula, the current in
the lead-p (either real or virtual lead) with spin index $\sigma$
can be expressed as
\begin{eqnarray}
 J_{p\sigma}=(e/\hbar)\sum_{q\neq p} T_{pq}^{\sigma }(V_{p\sigma}-V_{q\sigma}),
\end{eqnarray}
where $V_{p\sigma}$ is the spin-dependent bias in the lead p.
$T_{pq}^{\sigma}=Tr[{\bf \Gamma_{p\sigma}G^r\Gamma_{q\sigma}G^a }]$
is the transmission coefficient from the lead-q to p with spin
$\sigma$, where the linewidth functions ${\bf
\Gamma_{p\sigma}}=i[{\bf \Sigma_{p\sigma}^r}-{\bf
\Sigma_{p\sigma}^{r+}}]$, the Green function ${\bf G}^r=[{\bf
G}^a]^\dagger=[E_F{\bf I}-{\bf H}_{cen}-\sum_{p\sigma}
{\bf\Sigma}_{p\sigma}^r]^{-1}$, ${\bf H}_{cen}$ is the Hamiltonian
in the central region, and ${\bf \Sigma_{p}^r}$ is the retarded
self-energy due to the coupling to the lead-p.\cite{ref13} For the
real lead-p ($p=1,2,3,4,5,6$), the self-energy ${\bf \Sigma_{p}^r}$
can be calculated numerically. For the virtual leads, $ {\bf
\Sigma_{p}^r} =-i\Gamma/2$ and $\Gamma$ is the dephasing strength.

In our simulations, a small external bias is applied between the
lead-1 and lead-4 with
$V_{1\uparrow}=V_{1\downarrow}=-V_{4\uparrow}=-V_{4\downarrow}=V$,
which drives a current $I_{14}=I_1=-I_4$ flowing along the
longitudinal direction. For normal dephasing, electrons only lose
the phase memory while maintain the spin memory by going into and
then coming back from the virtual leads. Thus, for each virtual
lead-${\bf i}$ the currents have the constraint that $J_{{\bf
i}\uparrow}=J_{{\bf i}\downarrow}=0$, and $V_{{\bf i}\uparrow}$ is
usually not equal to $V_{{\bf i}\downarrow}$. But for spin
dephasing, electrons can lose both phase and spin memories, so one
has $V_{{\bf i}\uparrow}=V_{{\bf i}\downarrow}$ and $J_{{\bf
i}\uparrow}+J_{{\bf i}\downarrow}=0$ for each virtual lead-${\bf
i}$. In the recent experiment,\cite{ref9} the four transverse real
leads are the voltage probes, so $J_{p\uparrow}+J_{p\downarrow} = 0$
and $V_{p\uparrow}=V_{p\downarrow} \equiv V_p$ for $p=2$, 3, 5, and
6. Combining Eq.(2) together with all boundary conditions for the
real and virtual leads, the voltage $V_{p\sigma}$ and current
$J_{p\sigma}$ in each real lead can be obtained. Then the
longitudinal resistance $R_{14,23} \equiv V_{23}/I_{14}
=(V_2-V_3)/(I_{1\uparrow}+I_{1\downarrow})$ and $I_{2s}/I_{14}=
(I_{2\uparrow}-I_{2\downarrow})/I_{14}$ can be calculated and will
be presented next. Here $R_{14,23}$ is the measured quantity in the
recent experiment.\cite{ref9} In addition, we also consider the case
in which the four transverse leads are taken as spin-bias probes
with their currents $J_{p\uparrow} = J_{p\downarrow} = 0$. In this
case, we define a new spin Hall resistance $R_s \equiv
(V_{i\uparrow}-V_{i\downarrow})/I_{14}$ ($i$ can be any transverse
lead, for instance $i=2$), and its result will be shown in this
study as well.

\begin{figure}%[tbp]
\includegraphics[width=9cm,totalheight=6cm]{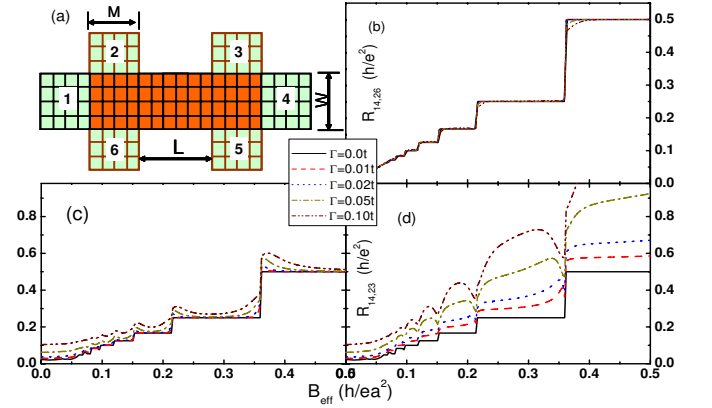}
\caption{(color online) (a) Schematic diagram for a six-terminal
Hall bar sample, the gray (or red) area is the central region
containing dephasing. (b) The Hall resistance $R_{14,26}$ vs. the
magnetic field $B$ for different dephasing strengths $\Gamma$. (c)
and (d) illustrate the longitudinal resistance $R_{14,23}$ vs $
B_{eff}$ in the presence of normal and spin dephasings,
respectively. The parameters are $M=24a$, $L=32a$ and $W=32a$. }
\end{figure}

In the numerical calculations, we take the hopping matrix element
$t=1$ as the energy unit. The Fermi energy is selected at $E_F=-3t$
which is near the energy-band bottom $-4t$. Since the flux in a unit
lattice is $\phi=1$ when the efficient magnetic field ${
B_{eff}}=h/(ea^2)$, $h/(ea^2)$ was taken as the unit of ${
B_{eff}}$. The dephasing strength is described by the parameter
$\Gamma$, which is directly related to the phase coherence length
$L_\phi$\cite{ref16}, an experimental observable parameter. Fig.2b
and 2c show the relation of $L_\phi$ vs $\Gamma$. With increasing
$\Gamma$, $L_\phi$ decreases rapidly and monotonically for either
normal or spin dephasing. To test out our model, we first
investigate the effect of dephasing on the integer QHE [i.e.
$\eta(\sigma)=1$ in the Hamiltonian (1)]. As shown in Fig.1b, the
quantized Hall resistance plateaus of $R_{14,26}$ in QHE is hardly
affected by either dephasing, in agreement with previous
experimental and theoretical results.\cite{ref16}

\begin{figure}%[tbp]
\includegraphics[width=8.5cm,totalheight=4cm]{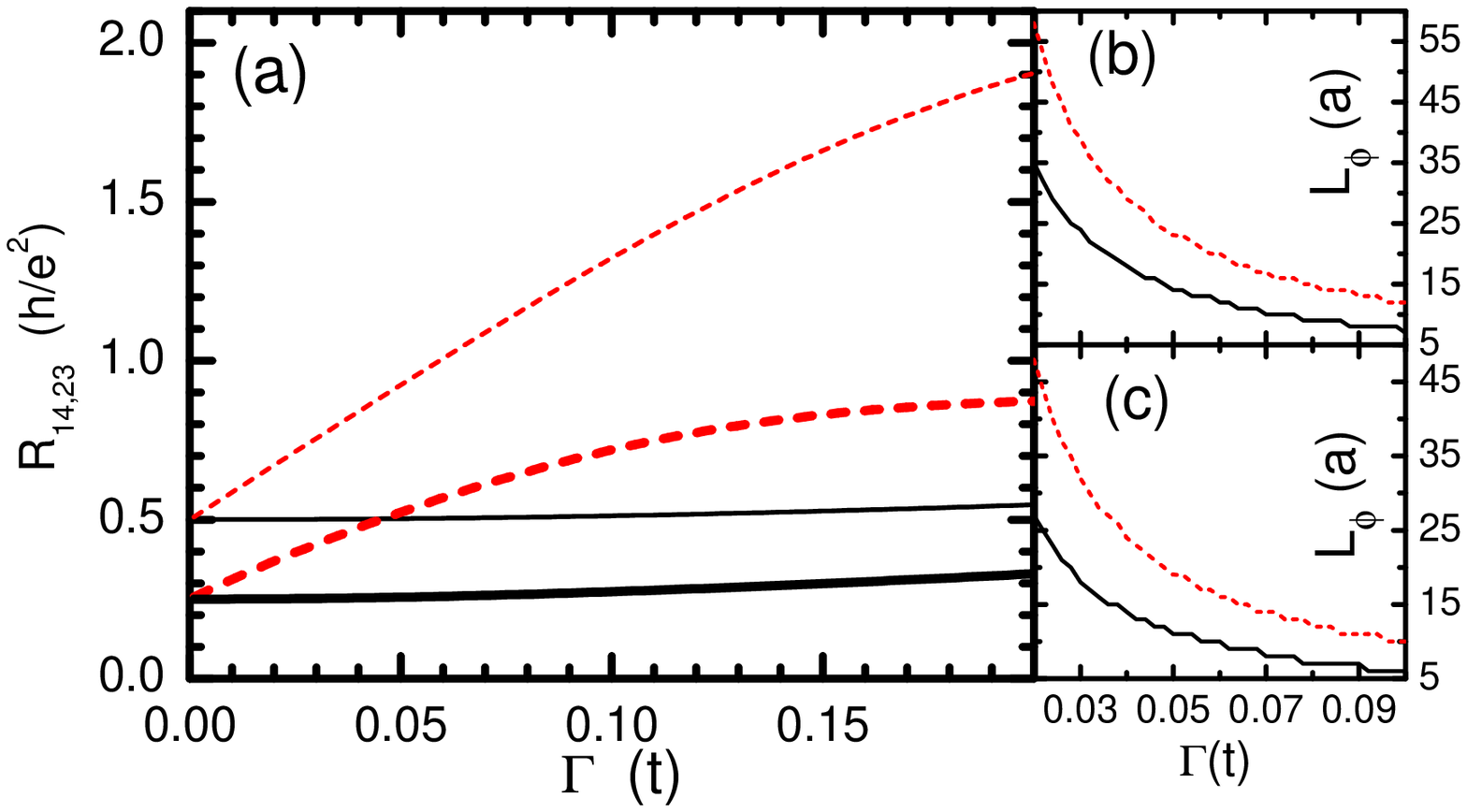}
\caption{ (color online) (a) shows $R_{14,23}$ vs. $\Gamma$ with
$L=32a$, $W=32a$, $M=24a$, and ${ B_{eff}}=0.5$ (thin curves) and
$0.3$ (thick curves). (b) and (c) show $L_\phi$ vs. $\Gamma$ at ${
B_{eff}}=0.5$ (b) and $0.3$ (c) with $W=32a$. The solid and dotted
curves in (a,b,c) are for the normal and spin dephasings.}
\end{figure}

Now, we present our numerical results of the dephasing effect on
QSHE. Fig.1c and 1d show the longitudinal resistance $R_{14,23}$
versus ${ B_{eff}}$ for the normal and spin dephasings with
different dephasing strength $\Gamma$. In the absence of dephasing
(solid lines), $R_{14,23}$ exhibits perfect quantum plateaus at
$h/2\nu e^2$ ($\nu=1,2,3,...$). Note that in the experiment of
Ref.\cite{ref9}, only one plateau was observed since only one edge
channel is there for each spin component at a given edge. In our
theoretical model of Eq.(1), one allows multi-channels. Thus, the
experimental situation corresponds to the highest plateau at
$h/2e^2$ with $\nu=1$ in our model. In the presence of dephasing,
the quantum plateaus of $R_{14,23}$ behave quite differently
depending on the type of dephasing. With normal dephasing, the
plateau structure remains and $R_{14,23}$ changes only slightly in
between the plateaus. For normal dephasing, temperature causes the
dephasing broadening in $\Gamma$, thus, this shows that QSHE is
insensitive to $T$ at low temperatures. However, from Fig.(1d), one
sees that with spin dephasing, $R_{14,23}$ increases significantly
even with small $\Gamma$.

Next, we investigate the effect of dephases in more detail. Fig.2a
shows $R_{14,23}$ versus dephasing strength $\Gamma$ for fixed ${
B_{eff}}=0.5$ and $0.3$, which are at the centers of the 1st and 2nd
plateaus. For the normal dephasing, the increase of $R_{14,23}$ is
extremely slow with increasing $\Gamma$. For example, at
$\Gamma=0.2t$, the sample size is about one order larger than the
phase coherence length $L_\phi$, but $R_{14,23}$ is only increased
by less $4\%$. In contrast, for the spin dephasing, $R_{14,23}$
increases rapidly with increasing $\Gamma$. For example, even for a
small $\Gamma=0.02t$, in which the sample length is shorter than
$L_\phi$, $R_{14,23}$ is increased by about $16\%$. For normal
dephasing, the carriers maintain their spin memories, and
backscattering only occurs when a carrier is scattered from one
boundary to the opposite one. So the backscattering is very weak
except when the Fermi energy is near a Landau level center, and the
quantum plateaus of $R_{14,23}$ can survive even with very large
normal dephasing. But for the spin dephasing, the spin of a carrier
can be flipped, and the backscattering occurs on each boundary. So
the longitudinal resistance $R_{14,23}$ is strongly affected by the
spin dephasing. In a real experimental sample, the spin dephasing
always exists to some degree, due to magnetic impurities, nuclear
spin fluctuations, etc., thus, the quantum plateaus of longitudinal
resistance of QSHE only survive in mesoscopic samples. This explains
why the quantum plateau was not observed in samples with large
lengths.\cite{ref9}

\begin{figure}%[tbp]
\includegraphics[width=8.5cm,totalheight=4cm]{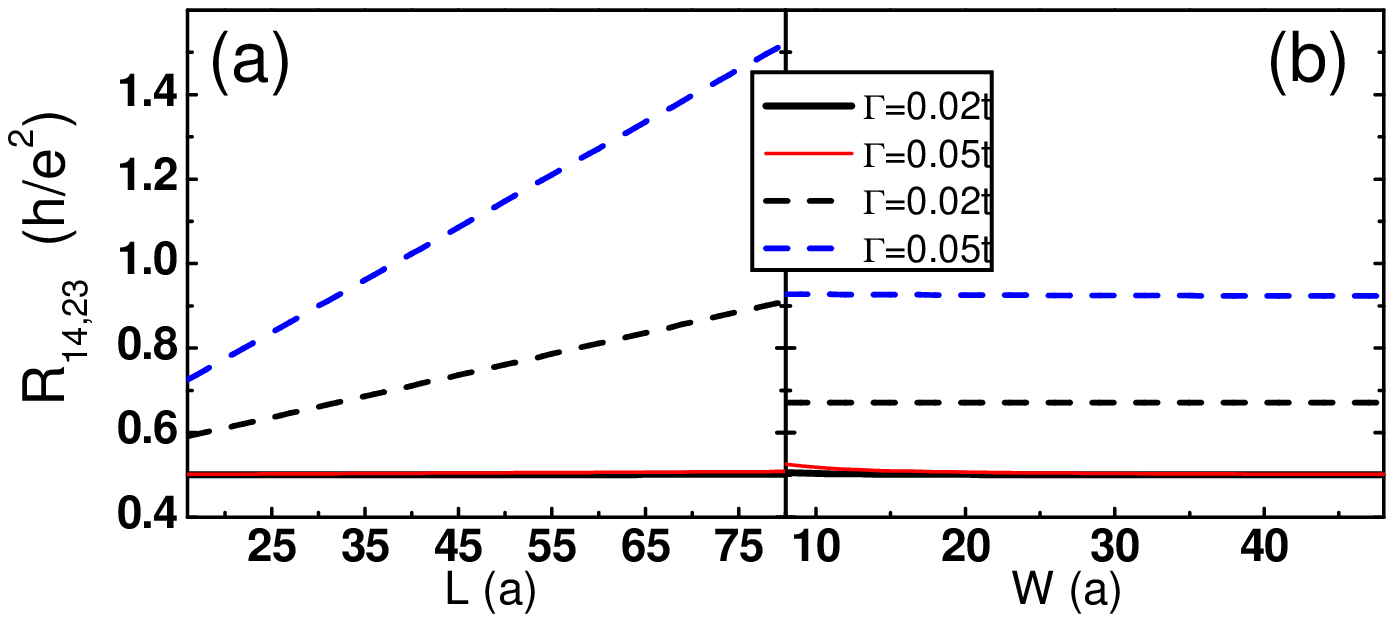}
\caption{(color online) (a) and (b) show $R_{14,23}$ vs. the sample
length $L$ with $W=32a$ (a) and the width $W$ (b) with $L=32a$ (b),
$M=24a$, and $B_{eff}=0.5$. The solid and dash lines are for the
normal and spin dephasings, respectively.}
\end{figure}

\begin{figure}%[t]
\includegraphics[width=8cm,totalheight=4cm]{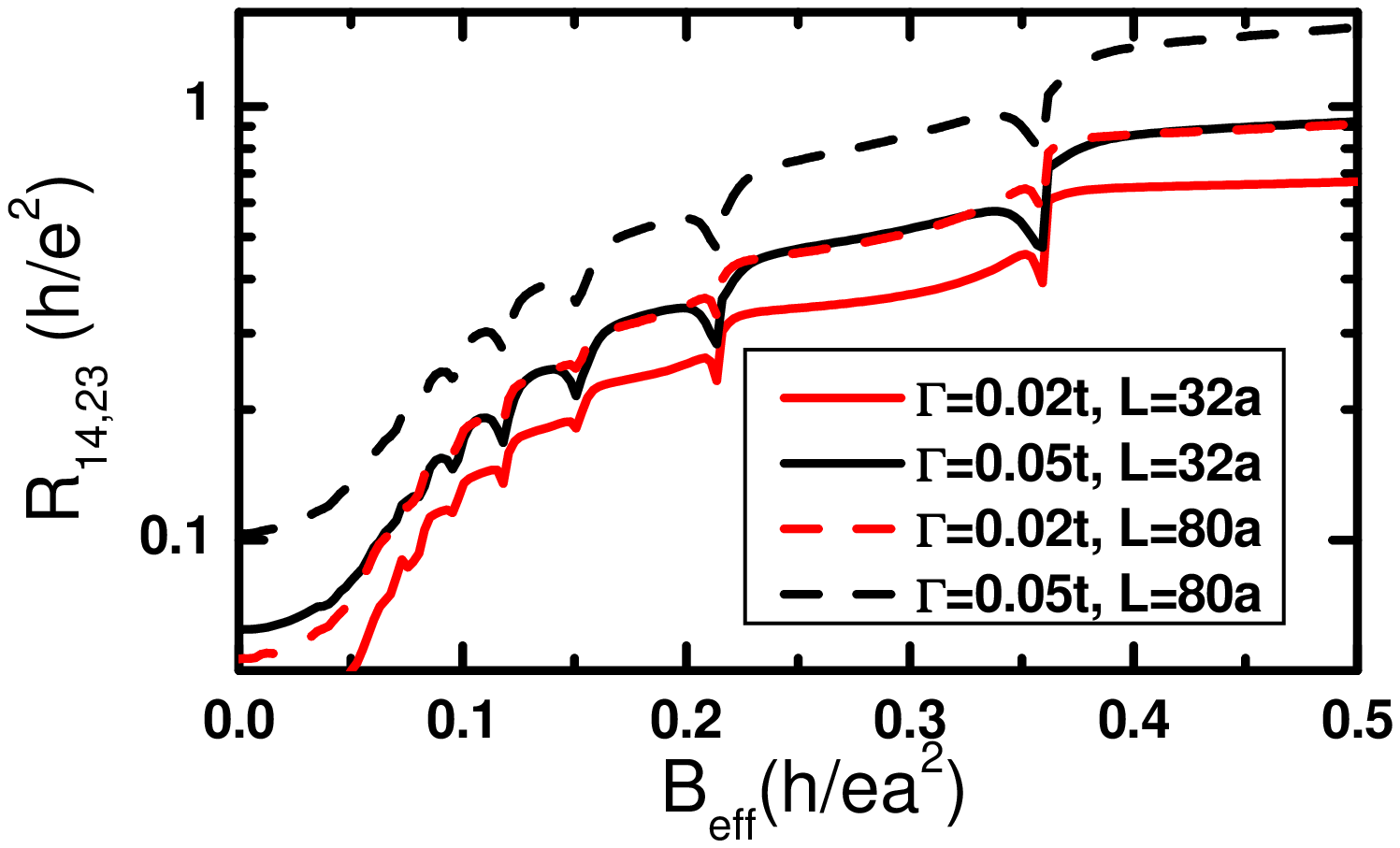}
\caption{(color online)  $R_{14,23}$ vs. ${ B_{eff}}$ for different
sample lengths $L$ and spin-dephasing strengths $\Gamma$ with
$W=32a$ and $M=24a$.}
\end{figure}

In Fig.3, $R_{14,23}$ dependence on the system sizes is studied. For
normal dephasing, the plateau of $R_{14,23}$ stays at the quantized
value regardless of the sample length $L$ or the width $W$ since the
backscattering is weak in all cases except for very small $W$. On
the other hand, for spin dephasing, $R_{14,23}$ is almost
independent of the width $W$ but is linearly increasing with the
length $L$ since the backscattering is stronger with larger $L$. In
Fig.4, we plot $R_{14,23}$ in logarithmic scale as done in the
experimental figures.\cite{ref9} Similar to the experimental plots,
$R_{14,23}$ approximatively shows the plateau characteristics
regardless of the dephasing strength $\Gamma$, although the plateau
values may well exceed the idealized quantized-values of $h/2\nu
e^2$ in the absence of spin dephasing. Combining all the results
from Figs.1-4, we qualitatively explain the experimental findings on
the behavior of the longitudinal resistance $R_{14,23}$ and its
dependence on temperature, sample length and sample
width.\cite{ref9}

\begin{figure}%[t]
\includegraphics[width=8.5cm,totalheight=7cm]{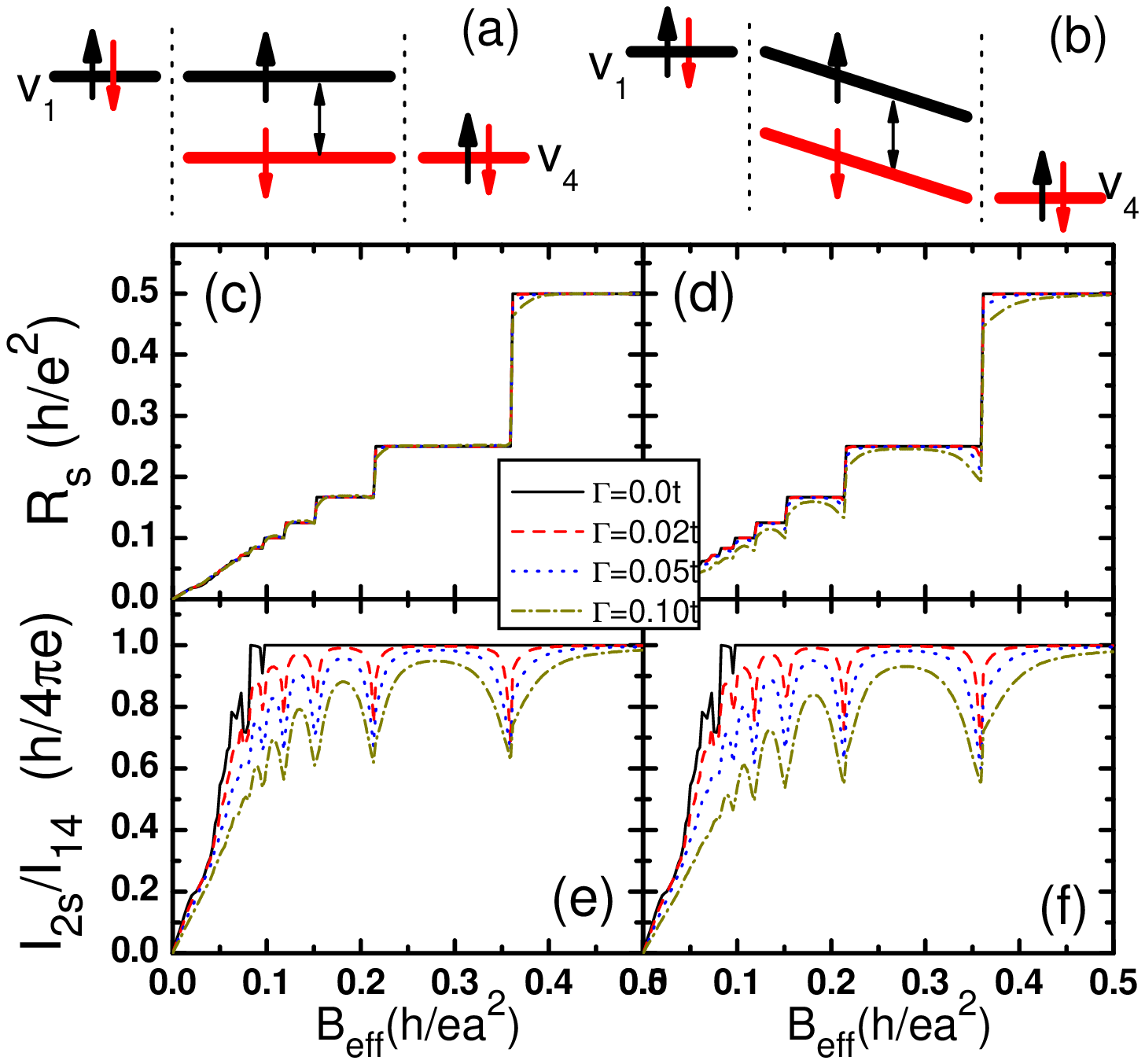}
\caption{(color online) (a) and (b) are the schematic diagrams for
the chemical potential along a given boundary with (a) and without
(b) spin dephasing. (c) and (d) plot the spin Hall resistance $R_s$
vs. ${B_{eff}}$. (e) and (f) illustrate the function $I_{2s}/I_{14}$
vs. ${ B_{eff}}$. (c) and (e) are for the normal dephasing case, and
(d) and (f) are for the spin dephase case. The parameters are
$M=24a,L=32a$ and $W=32a$.}
\end{figure}

Up to now, we find that the quantum plateaus of longitudinal
resistance $R_{14,23}$ survive only for mesoscopic samples due to
spin dephasing. Whether there exists an observable physical quantity
in macroscopic QSHE samples? Or more importantly, is there an
observable quantity better reflect the topological nature of the
QSHE than $R_{14,23}$. We find that a new spin Hall resistance $R_s
\equiv (V_{i\uparrow}-V_{i\downarrow})/I_{14}$ ($i$ can be any
transverse lead, for instance $i=2$) can fulfill the purpose.
Figs.5c and 5d show the spin Hall resistance $R_s$, and it exhibits
the quantum plateaus at $h/2\nu e^2$ even with strong normal or spin
dephasing. For example, at $\Gamma=0.1t$, the sample length exceeds
$L_{\phi}$ by one order of magnitude, the plateaus of $R_s$ still
stay at the quantized values. This means that these plateaus will be
visible in macroscopic QSHE samples. The robustness of $R_s$ against
either dephasing is similar to what appeared in the Hall resistance
plateaus in the conventional QHE (see Fig.1(b)). Let us explain the
origin of the story with the aid of Figs.5a,b, in which the chemical
potential along a given boundary is shown. In the left and right
leads, the chemical potentials are always spin-independent with
$V_{1\sigma}=-V_{4\sigma}=V$. In the central region, the chemical
potential $V_{c\sigma}$ is spin-dependent. Without spin dephasing,
$V_{c\uparrow}=V$ and $V_{c\downarrow}=-V$ (see Fig.5a) since the
spin-up electrons flow to the right while the spin-down electrons to
the left. In the presence of spin dephasing, the chemical potential
$V_{c\sigma}$ descends along the longitudinal direction (see
Fig.5b). But in order to keep the current $I_{14}$ to be a constant,
$V_{c\uparrow}-V_{c\downarrow}$ needs to be also unchanged
regardless of the positions along the sample and the spin dephasing
strength $\Gamma$ since the current $I_{14}$ is carried by the edge
states between $V_{c\uparrow}$ and $V_{c\downarrow}$. Therefore, the
plateaus of the spin Hall resistance
$R_s=(V_{2\uparrow}-V_{2\downarrow})/I_{14}$ will stay unchanged
even with strong spin-dephasing (i.e. in macroscopic samples). In
addition, due to constant nature of $R_s$, the ratio of the
transverse spin current to the longitudinal current(e.g.
$I_{2s}/I_{14}$, see Fig.5e,f) and the spin accumulation on the
boundary can also survive in the presence of strong normal and spin
dephasings. Note that the spin accumulations and the difference of
$V_{2\uparrow}-V_{2\downarrow}$ have been measured in recent
experiments.\cite{ref17,ref18}

In summary, the effect of dephasing on QSHE is studied. We find that
the quantum plateaus of the longitudinal resistance $R_{14,23}$ are
insensitive to the normal dephasing, but are severely affected by
the spin dephasing, so that these quantum plateaus exist only in
mesoscopic samples. This result explains why the quantum plateaus of
$R_{14,23}$ are only observed in small-size samples in the recent
experiment.\cite{ref9}. The dephasing effect also provides the
understanding of observed dependence of $R_{14,23}$ on temperature,
sample length and sample width. In addition, we find a new spin Hall
resistance that also exhibits quantum plateaus. In particular, these
plateaus stay at quantized values in macroscopic samples and better
reflect the topological nature of QSHE.

This paper is supported by NSF-China under Grants Nos. 10525418 and
10734110, by US-DOE under Grants No. DE-FG02- 04ER46124 and US-NSF.

\end{document}